\pdfoutput=1

\documentclass[aps,prb,twocolumn]{revtex4}
\usepackage{makeidx}
\usepackage{amsmath,amssymb,amsfonts,amsthm}
\usepackage{graphicx,bm}

\begin{document}

\title{Effect of doping and oxygen vacancies on the octahedral tilt
transitions in the BaCeO$_{3}$ perovskite}
\date{}
\author{F. Cordero,$^{1}$ F. Trequattrini,$^{2}$ F. Deganello,$^{3}$ V. La
Parola,$^{3}$ E. Roncari$^{4}$ and A. Sanson$^{4}$}
\affiliation{$^{1}$ CNR-ISC, Istituto dei Sistemi Complessi, Area della Ricerca di Roma -
Tor Vergata,\\
Via del Fosso del Cavaliere 100, I-00133 Roma, Italy}
\affiliation{$^{2}$ Dip. Fisica, Universit\`{a} di Roma "La Sapienza", P.le A. Moro, 5
I-00184 Roma, Italy}
\affiliation{$^{3}$ CNR-ISMN, Istituto per lo Studio dei Materiali Nanostrutturati,\\
Via Ugo La Malfa 153, I-90146 Palermo, Italy}
\affiliation{$^{4}$ CNR-ISTEC, Istituto di Scienza e Tecnologia dei Materiali Ceramici,\\
Via Granarolo 64, I-48018 Faenza, Italy}

\begin{abstract}
We present a systematic study of the effect of Y doping and
hydration level on the structural transformations of BaCeO$_{3}$
based on anelastic spectroscopy experiments. The temperature of the
intermediate transformation between rhombohedral and orthorhombic
\textit{Imma} phases rises with increasing the molar fraction $x$ of
Y roughly as $\left( 500~\text{K}\right) \times x$ in the hydrated
state, and is depressed of more than twice that amount after
complete dehydration. This is explained in terms of the effect of
doping on the average (Ce/Y)-O and Ba-O bond lengths, and of lattice
relaxation from O vacancies. The different behavior of the
transition to the lower temperature \textit{Pnma} orthorhombic phase
is tentatively explained in terms of progressive flattening of the
effective shape of the OH$^{-}$ ion and ordering of the O vacancies
during cooling.
\end{abstract}

\maketitle


\section{Introduction}

The BaCeO$_{3}$ perovskite undergoes three phase transformations starting
from the high temperature cubic (C) phase: to rhombohedral (R) at $%
T_{1}=1170 $~K, to orthorhombic \textit{Imma} (O1) at $T_{2}=$ 670~K and to
orthorhombic \textit{Pnma} (O2)\ at $T_{3}=$ 563~K, as determined by neutron
diffraction\cite{Kni01} and by combined differential scanning calorimetry,
dilatometry and X-ray diffraction.\cite{OHS09} The sequence of phase
transformations and the various structures are well characterized in the
undoped state of BaCeO$_{3}$, and even a quantitative description of the
spontaneous strains by means of the Landau expansion of the free energy has
been presented.\cite{Dar96} Instead, the situation is confused when a
trivalent dopant, e.g. Y$^{3+}$, is substituted into the Ce$^{4+}$ place in
order to make the material a protonic conductor. The understanding of the
influence of doping on the phase transitions in the perovskite ionic
conductors is not only of academic interest. In fact, the occurrence of
phase transformations, especially if accompanied by ordering of the mobile
ionic species, protons and O vacancies, is closely related to the mobility
of such ions and to the durability of the material in applications like fuel
cells or membranes for gas separation.\cite{Kre97,BHG02,NIS07,CMT09}
Although various indications have been reported that the transition
temperatures in BaCeO$_{3}$ depend on doping and on the hydration state,\cite%
{BEK98,Kni01,YY03,SVA93,Kni99,TLR00,KMJ05} no systematic study and
analysis has appeared yet. According to Raman spectroscopy
measurements on variously doped BaCeO$_{3}$, the room temperature
structure changes to the more symmetric tetragonal and cubic phases
with increasing Nd$^{3+}$ substitution,\cite{SVA93} but a subsequent
neutron diffraction experiment excludes any significant influence on
the room temperature orthorhombic structure from Nd
doping.\cite{Kni99} For BaCe$_{1-x}$Y$_{x}$O$_{3-\delta }$ (BCY), a
change from the O2 to the R structure at room temperature was found
at $x\geq 0.2$ with neutron diffraction,\cite{TLR00} whereas a later
x-ray diffraction study did not show such a transition to
rhombohedral at room temperature, but rather
impurity phases arising from a more limited Y solubility range.\cite%
{KMJ05,GLD07} More recently, anelastic spectroscopy measurements on BCY
showed that passing from the hydrated to the outgassed state with $x=0.1$
lowers the temperature of the O1-R transition by as much as 250~K.\cite{137}
Here we present a more extensive study of the effect of doping and O
vacancies (V$_{\text{O}}$) on the phase transitions in BCY, again based on
anelastic measurements. An interpretation of the observations is proposed,
assuming that the driving force for the octahedral tilting instabilities is
the mismatch between too long A-O and too short B-O bonds, as usual for ABO$%
_{3}$ perovskites; a minimal model is adopted for the changes with
doping and hydration level of the tolerance factor and of the
lattice relaxation due to V$_{\text{O}}$.

\section{Experimental}

The samples of BaCe$_{1-x}$Y$_{x}$O$_{3}$ with $x=$ 0, 0.02, 0.1,
0.15, 0.3 were prepared as already described,\cite{132} with
starting powders obtained by auto-combustion synthesis,\cite{DMD08}
followed by crystallization in air at 1273~K for 5~h. No oxide
impurity phases were detected by x-ray diffraction (XRD) after
synthesis for $x\leq 0.15$. The sample with nominal $x=0.3$\emph{\
}was not monophasic, since the solubility limit of Y
in BCY is lower than 0.3. In fact, impurity phases are detected by XRD for $%
x\geq 0.2$\cite{KMJ05} and by EXAFS for $x\geq 0.17$.\cite{GLD07} We did not
determine the exact concentration of Y\ in solid solution in the $x=0.3$
sample and in what follows we will set this value to $x=0.2$. The powders
were first uniaxially pressed at 50~MPa and then isostatically pressed at
200~MPa obtaining $60\times 7\times 6$~mm ingots, which were sintered at
1773~K for 10~h. The samples were cut as thin bars about 1~mm thick and $%
\sim 4$~cm long. In order to make them conducting for the anelastic
experiments, their faces where covered with Ag paint, or Pt paint when
temperatures higher than 900~K had to be reached. We tried with SPI-Chem
Conductive Platinum Paint consolidated at 1270~K, or with 1000~\AA\ of Pt
magnetron sputtered directly on the sample surface. Unfortunately, none of
these electrodes resisted the anelastic measurements in vacuum $<10^{-5}$%
~mbar in the temperature range 1000-1300~K, since in all cases they
evaporated away. This fact rendered the measurements at $>1000$~K difficult,
and we could not obtain extensive and reliable data on the R-C transition
near 1200~K.

Hydration was achieved by maintaining the samples for 1-2~h at 793~K in a
static atmosphere of $50-100$~mbar H$_{2}$O, followed by slow cooling, while
outgassing was achieved in vacuum $<10^{-5}$~mbar up to 1000~K or during the
anelastic experiments. The resulting variations of the gaseous contents were
monitored from the change of weight. The reaction of equilibrium of the O\
deficient perovskite with water vapor is H$_{2}$O$+$V$_{\text{O}}^{\bullet
\bullet }+$O$_{\text{O}}^{\times }\longleftrightarrow 2$OH$_{\text{O}%
}^{\bullet }$,\cite{Kre97} where a molecule of water fills one V$_{\text{O}}$
and provides two protons that may diffuse among O$^{2-}$, to which are bound
as peroxide ions (OH)$^{-}$. The superscript dots represent excess $+e$
charges of the species with respect to the perfect lattice in Kr\"{o}%
ger-Vink notation. According to this reaction,\ the concentration of V$_{%
\text{O}}$ in BaCe$_{1-x}$Y$_{x}$O$_{3-\delta }$H$_{y}$ can vary within $%
\delta \leq x/2$ and concomitantly the concentration $y$ of protons
within $y\leq x$. It was found that the maximum possible hydration
was $\sim 15\%$ lower than the theoretical maximum of $y=x$; this is
usual in the doped cerate and zirconate perovskites, and can be due
to partial occupation of the Ba sites by the trivalent dopants or to
other defects.

The Young's modulus $E$ was measured by electrostatically exciting
the flexural modes of the bars suspended in vacuum on thin
thermocouple wires in correspondence with the nodal lines. Besides
the 1st flexural mode, the 3rd and sometimes the 5th modes, with
frequencies $5.4$ and $13.3$ times higher, could be measured during
the same run; the frequency of the fundamental mode was $\omega
/2\pi \simeq 1.5-3$\ kHz, depending on the sample shape and state.
The data will be presented as real part $s^{\prime }$ of the elastic
compliance $s\left( \omega ,T\right) =s^{\prime }-is^{\prime \prime }=$ $%
E^{-1}$, referred to its extrapolation $s_{0}$ to 0~K, and elastic energy
loss coefficient $Q^{-1}=$\ $s^{\prime \prime }/s^{\prime }$. The first is
proportional to the square of the sample resonance frequency, $s^{\prime
}\left( T\right) \propto \omega ^{2}\left( T\right) $, and presents peaks or
steps at the structural phase transformations; the latter was measured from
the decay of the free oscillations or from the width of the resonance peak,
and presents peaks due to the relaxational motion of point and extended
defects\cite{NB72} (V$_{\text{O}}$, protons and their complexes with
dopants, twin walls, etc.). For an elementary relaxation process it is\cite%
{NB72}
\begin{equation}
\delta s\left( \omega ,T\right) =\frac{\Delta }{T}\frac{1}{1+i\omega \tau }~,
\label{Debye}
\end{equation}%
\ with maximum at the temperatures at which the defect relaxation time $\tau
\sim \omega ^{-1}$. Since $\tau \left( T\right) $ is a decreasing function
of temperature, usually according to the Arrhenius law, the temperature of a
thermally activated peak increases with frequency. The peaks due to the
hopping of V$_{\text{O}}$ and to the reorientation of protons around Y
dopants have already been identified,\cite{132,137} and allow one to monitor
the concentrations of such defects and to study their dynamics.

\section{Results}

\begin{figure}[tbp]
\includegraphics[width=8.5 cm]{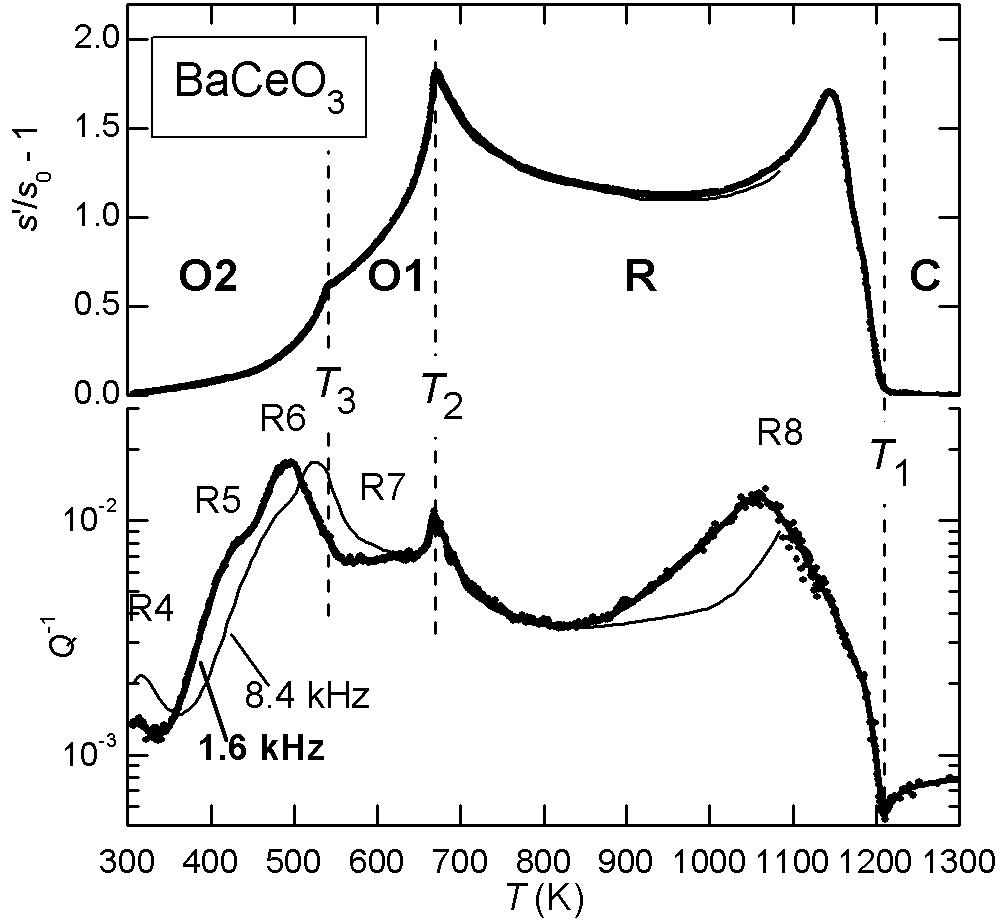}
\caption{Real part of the elastic
compliance (upper panel) and elastic energy loss coefficient (lower panel)
of undoped BaCeO$_{3}$, measured at 1.6 and 8.4~kHz.}
\label{fig1}
\end{figure}

Figure \ref{fig1} presents the anelastic spectrum of a sample of undoped
BaCeO$_{3}$ measured at two frequencies: 1.6 and 8.4~kHz. The real parts $%
s^{\prime }$ are practically coincident at both frequencies and present
sharp steps or peaks in correspondence with the three phase transformations
at $T_{1}=1210$~K, $T_{2}=668$~K and $T_{3}=540$~K. These temperatures are
close to those determined by neutron diffraction\cite{Kni01} and define the
temperature ranges of the cubic, rhombohedral and two orthorhombic phases.
We identify $T_{1}$ of the R-C transformation with the temperature of the
kink between almost flat and sharply rising compliance, rather than with the
peak at a temperature 65~K lower. This also coincides with a sharp dip in
the absorption. Such a dip is rather anomalous, since usually one finds a
peak or more or less rounded step at the onset of a structural
transformation, but it clearly separates the rhombohedral region, with
absorption due to the motion of domain walls, from the cubic region without
appreciable anelastic losses.


The elastic energy loss coefficient, besides clear anomalies in
correspondence with the transitions, has five relaxation peaks in the $%
300-1300$~K temperature range, labeled R4-R8 because there are other
relaxation processes at lower temperatures (see Ref.
\onlinecite{132} and Fig. \ref{fig4} later on). The thermally
activated character of these processes is clear from the fact that
they are shifted to higher temperature at the higher frequency (see
Eq. (\ref{Debye})).


\begin{figure}[tbp]
\includegraphics[width=8.5 cm]{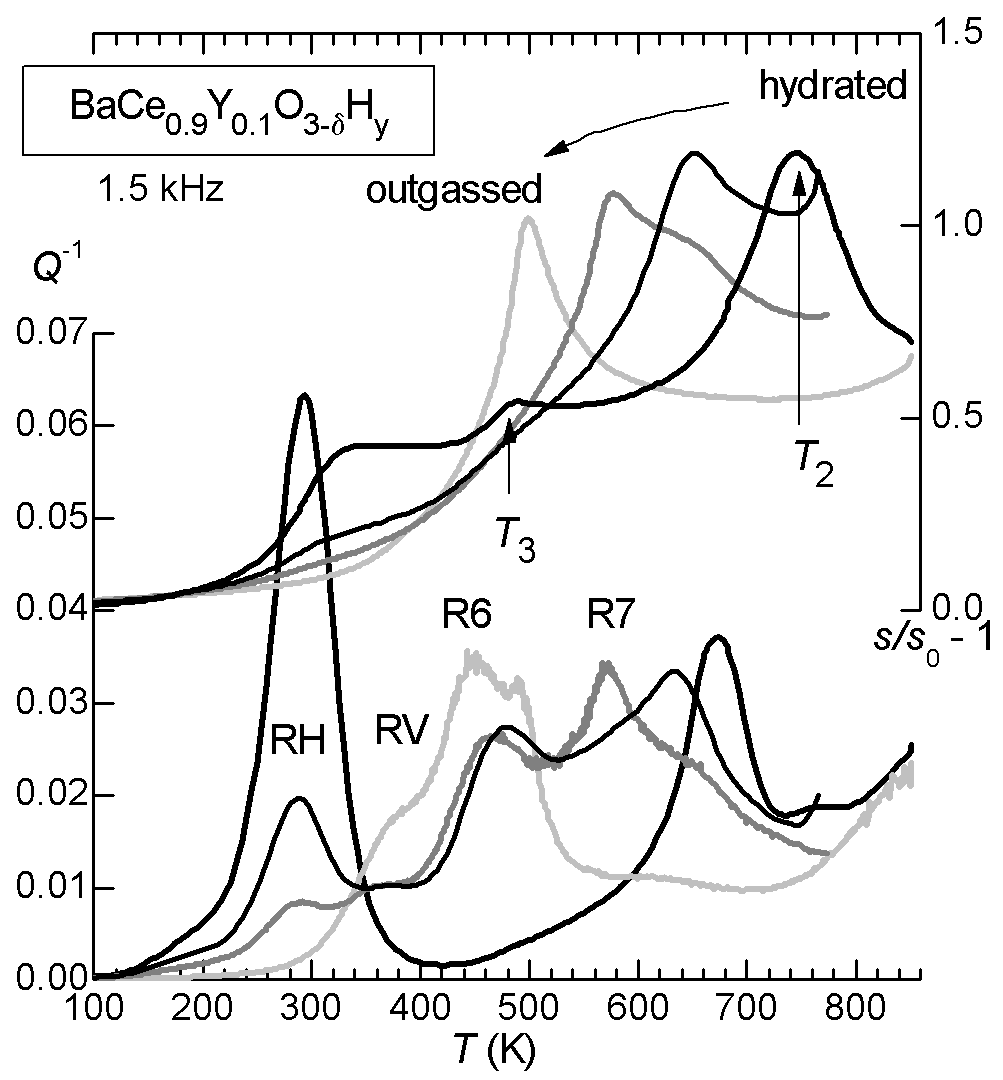}
\caption{Real part of the elastic
compliance (upper panel) and elastic energy loss coefficient (lower panel)
of BaCe$_{1-x}$Y$_{x}$O$_{3-\protect\delta }$H$_{y}$ with $x=0.1$, measured
at $\sim 1.5$~kHz at various hydration levels, from fully hydrated (black)
to fully outgassed (light grey).}
\label{fig2}
\end{figure}

Figure \ref{fig2} presents a series of anelastic spectra of a sample of BaCe$%
_{1-x}$Y$_{x}$O$_{3}$ with $x=0.10$ at various stages of hydration,
from fully hydrated (thick black lines)\ to fully outgassed (light
grey); these spectra have already been published in a preliminary
study\cite{137} of the effect of varying hydration on the structural
and elastic properties of BCY. The two $Q^{-1}\left( T\right) $
peaks at lower temperature are labeled as
RH and RV, since they are due to hopping of protons, likely around Y dopants,%
\cite{132} and of V$_{\text{O}}$, respectively. Their evolution allows us to
confirm that the sample passes from fully hydrated (RV is absent and RH
saturated)\ to fully outgassed (RH is absent and RV saturated). The presence
of the intermediate curves (only few of them are reported) allows us to
ascertain that indeed outgassing shifts the transition at $T_{2}$ of 250~K
to lower temperature, while the transition at $T_{3}$ is soon smeared and
masked by the presence of the former transition, but does not seem to shift
appreciably.


\begin{figure}[tbp]
\includegraphics[width=8.5 cm]{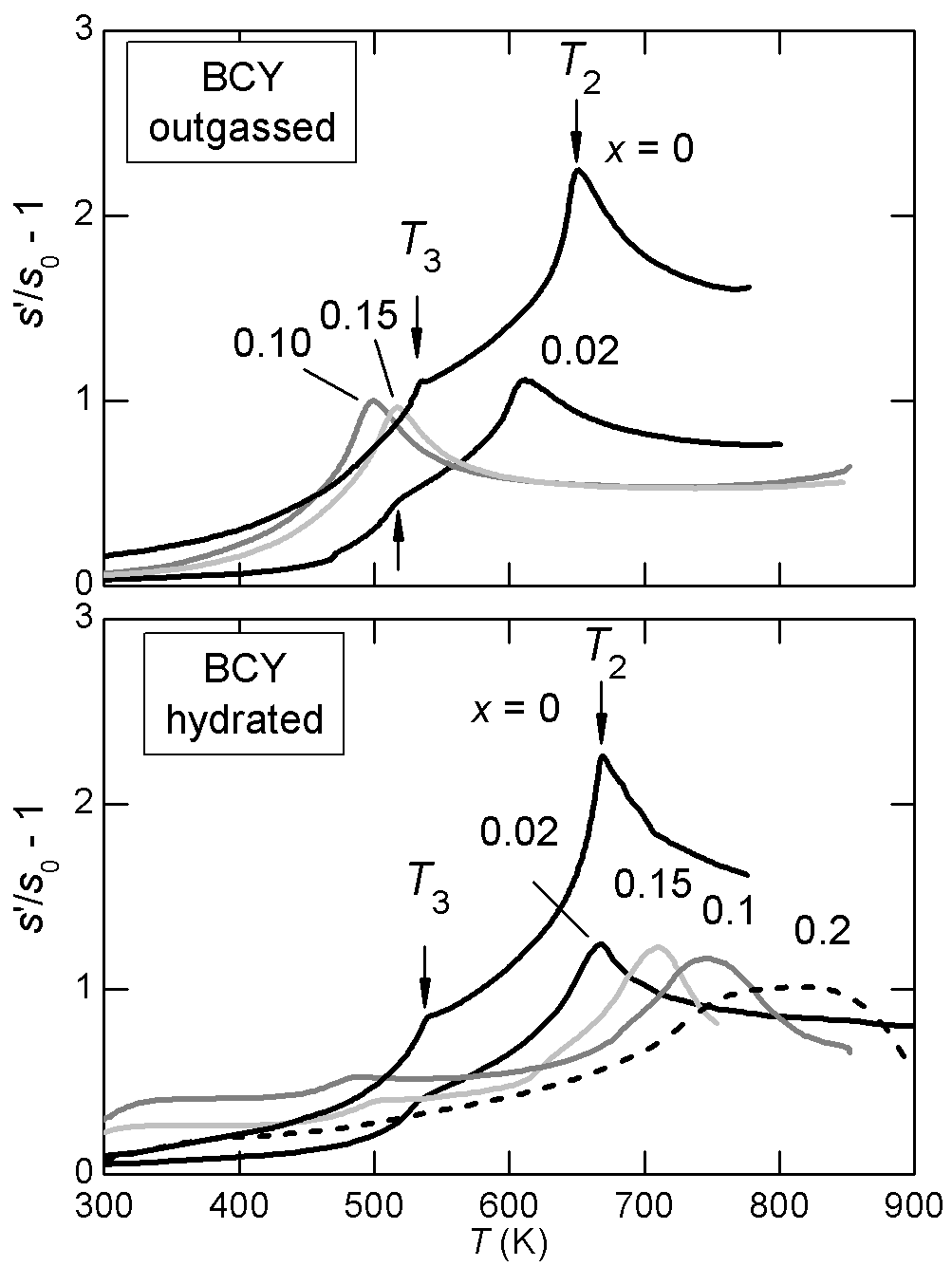}
\caption{Real part of the elastic
compliance of a series of samples with different doping $x$ measured in the
hydrated and outgassed states.}
\label{fig3}
\end{figure}

The effect of Y$^{3+}$\ doping on the phase transformations at $T_{2}$ and $%
T_{3}$ is shown in Fig. \ref{fig3}, where the elastic compliance curves are
plotted of samples having $x=$ 0, 0.02, 0.10, 0.15 and 0.2 in the fully
hydrated and fully outgassed states. There is no outgassed curve at $x=0.2$,
because the sample broke after the first measurement. With increasing
doping, and hence lattice disorder, there is progressive smearing of the
peaks at the transitions, so that $T_{3}$ becomes more and more difficult to
determine. This is especially true in the outgassed state, where the
transition at $T_{2}$, whose effects on the elastic compliance are
prevalent, shifts consistently to lower temperature and masks the effects of
the O2-O1 transition. Note that there is an inversion in the trend of the
anelastic spectra between $x=0.1$ and 0.15, as discussed later. The
transition temperatures deduced from these curves will be plotted in Fig. %
\ref{fig5}.


\begin{figure}[tbp]
\includegraphics[width=8.5 cm]{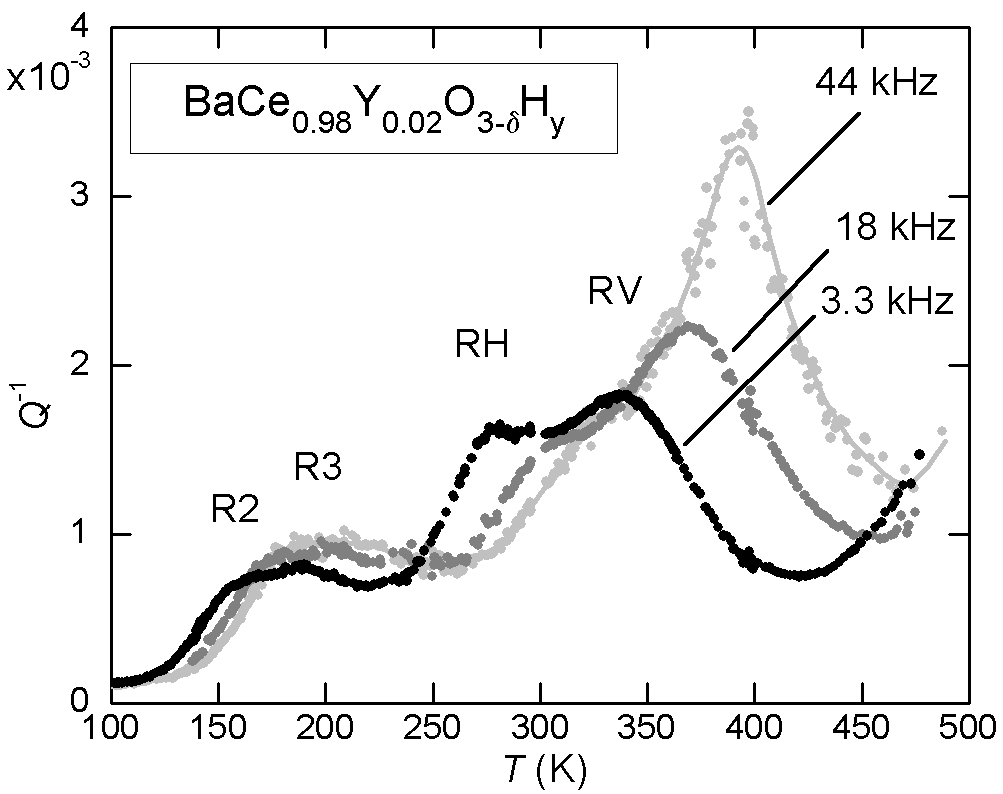}
\caption{Elastic energy loss coefficient
of BCY with $x=0.02$ in an intermediate hydration state, measured at three
frequencies.}
\label{fig4}
\end{figure}

We finally present an example of $Q^{-1}\left( T\right) $ curves measured at
three different frequencies, where it is particularly clear that the
dynamics of the V$_{\text{O}}$ is not simply that of independent defects,
which would give rise to a Debye relaxation, Eq. (\ref{Debye}), but seems to
have an important contribution from cooperative effects, possibly connected
with V$_{\text{O}}$ ordering in the O2 phase. The $Q^{-1}\left( \omega
,T\right) $ curves in Fig. \ref{fig4} are measured at 3.3, 18 and 44~kHz on
a sample with $x=0.02$ in an intermediate state of hydration where both V$_{%
\text{O}}$ and H are present. Among the various peaks, with the help of Fig. %
\ref{fig2} we recognize RH, probably due to reorientation of H about Y
dopants,\cite{132} while RV is certainly due to V$_{\text{O}}$, although it
is not yet determined whether trapping by Y has a role. The interesting
feature in Fig. \ref{fig4} is that the intensity of peak RV is a drastically
increasing function of temperature, instead of having the $1/T$ dependence
expected from Eq. (\ref{Debye}). The effect is not due to O loss during the
measurements in vacuum, since the curves at different frequencies are
measured during a same run, and we have abundant data showing that the $%
Q^{-1}\left( T\right) $ are perfectly reproducible until one does not exceed
500~K. We mention that also the relaxation peak R6 in Fig. \ref{fig2}
displays a similar behavior, although the divergence of its intensity on
approaching $T_{3}$ might be partially due to overlapping with the narrow
dissipation peak associated with the structural transition. We did not make
a thorough analysis of these complicated anelastic spectra, and cannot say
yet whether R6 involves V$_{\text{O}}$, twin walls or both; it certainly
appears that the motion of the V$_{\text{O}}$ has a high degree of
cooperativity below $T_{3}$.

\section{Discussion}

\begin{figure}[tbp]
\includegraphics[width=8.5 cm]{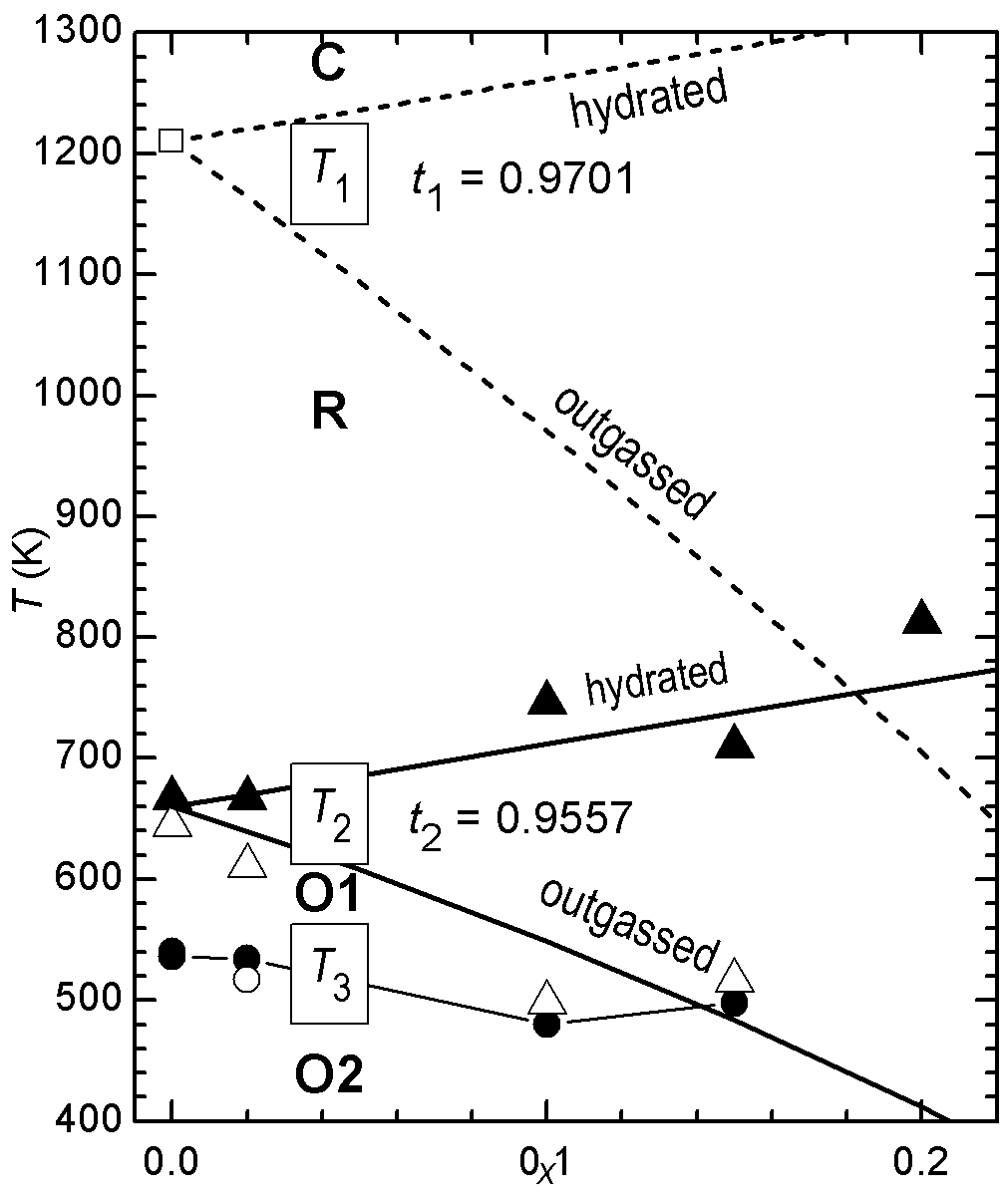}
\caption{Temperatures of the three
structural transformations of BCY in the hydrated (filled symbols) and
outgassed (open symbols) states: R-C square, O1-R triangles, O2-O1 circles.
The thick filled and dashed lines are calculated as explained in the text.}
\label{fig5}
\end{figure}

Figure \ref{fig5} shows the transition temperatures $T_{1}$, $T_{2}$ and $%
T_{3}$ plotted versus Y doping in both the fully hydrated and
outgassed states. The temperatures $T_{2}$ and $T_{3}$ are
determined from both the real parts $s^{\prime }\left( T\right) $ in
Fig. \ref{fig3} and the respective $Q^{-1}\left( T\right) $ curves
(not reported here), while $T_{1}$ is determined from Fig.
\ref{fig1}. The O1-R phase transformation is the better
characterized in the present measurements and exhibits the largest
dependence on doping: an increase of $T_{2}$ with doping in the
hydrated state and an even larger decrease in the outgassed state.
Notice that there
is a difference of nearly 30~K between outgassed and hydrated state even at $%
x=0$, which may be due to the contribution of electronic compensation or to
the presence of some defects, \textit{e.g.} from non perfect stoichiometry.
The transition at $T_{3}$ between the two orthorhombic structures is less
affected by doping and by O\ stoichiometry. Differently from that at $T_{2}$%
, it decreases slightly its temperature with doping in the hydrated state,
while only at $x=0.02$ it was possible to verify the lowering of the
transition temperature after outgassing, because of the masking effect of
the transition at $T_{2}$.


The R-C transformation has only one point for the undoped case,
because our results are only partial and preliminary, due to the
experimental difficulties explained in Section II. In addition, our
anelastic experiments are made in high vacuum, so that above 800~K
it is impossible to maintain the sample in the hydrated state, and
we can only measure reliably the temperature $T_{1}$ of BCY in the
outgassed state. Also in the literature there are no data on $T_{1}$
of doped BCY.

In what follows we will try to explain the fact that the temperature $%
T_{2}\left( x\right) $ increases with Y doping $x$ in the hydrated state,
and instead decreases with $x$ of an even larger amount in the outgassed
state. The uncertainty in the values of $T_{2}$ in Fig. \ref{fig5} is
smaller than the symbol size, and the fact that the points at $x=0.10$ and
0.15 do not follow a monotonic trend with doping is likely real and not an
experimental vagary. This anomaly can be put in relation with the
observation of a jump or extremum in the doping dependence of several
structural parameters of BaCeO$_{3-\delta }$ and SrCeO$_{3-\delta }$ at a
nominal concentration of V$_{\text{O}}$ $\delta \sim 1/16$,\cite{KMJ05}
corresponding to $\delta =2x=1/8$ in the fully outgassed state. This
phenomenon has been tentatively attributed to ordering of the V$_{\text{O}}$
commensurate with the lattice during the synthesis at high temperature,
hence with possible ordering of the cation dopants that would affect the
hydration properties also at lower temperature.\cite{KMJ05} We will ignore
this local inversion of the variation of $T_{2}\left( x\simeq 1/8\right) ,$
and only consider the positive average derivative of $T_{2}\left( x,\delta
=0\right) $ with respect to $x$, and negative derivative of $T_{2}\left(
x,\delta =x/2\right) $.

In searching for the relevant factors determining $T_{2}$, we note that the
O1-R transition involves tilting of the O octahedra, without the atomic
off-centering accompanying the ferroelectric transitions or additional
Jahn-Teller distortions, since neither Ce$^{4+}$ nor Y$^{3+}$ are
Jahn-Teller active. BCY is also inert from the magnetic point of view, so
that we conclude that the main driving force for the octahedra to tilt is
the mismatch between too long B-O bonds (B~= Ce/Y)\ and too short A-O bonds
(A~= Ba), as usual for ABO$_{3}$ perovskites.\cite{Goo04} The tendency of
perovskites to undergo tilting transitions is often expressed in terms of
the tolerance factor
\begin{equation}
t=\frac{r_{\mathrm{A}}+r_{\mathrm{O}}}{\sqrt{2}\left( r_{\mathrm{B}}+r_{%
\mathrm{O}}\right) }
\end{equation}%
where the mean ionic radii are the effective ones usually taken from
Shannon's tables.\cite{SP69} A value $t=1$ means that the ideal A-O and B-O
bond lengths, taken as the sums of the ideal ionic radii, perfectly match
the cubic structure, and therefore that the cubic phase should be stable; $%
t<1$ means that the B-O bond length is too large with respect to the A-O
one, and therefore that the octahedra tend to rotate in order to accommodate
the mismatch. The longer and weaker A-O bonds have larger thermal expansion
than the shorter and stronger B-O bonds. For this reason, perovskites with $%
t<1$ already at high temperature further decrease $t$ on cooling, until the
mismatch between too long B-O and too short A-O bonds is relived by a
tilting structural transformation. Usually, with decreasing $t$ below 1, one
finds first tilt patterns producing a more symmetric rhombohedral structure
and then the more distorted orthorhombic structures.\cite{Att01,Goo04} In
this respect, BaCeO$_{3}$ behaves normally, with the cubic phase
transforming into rhombohedral $R\overline{3}c$ and further into
orthorhombic O1 (\textit{I}$mma$) and O2 ($Pnma$). The tilt patterns in
Glazer's notation\cite{Gla72,Kni01} are respectively $a^{-}a^{-}a^{-}$, $%
a^{0}b^{-}b^{-}$ and $a^{+}b^{-}b^{-}$, and the anomaly in the sequence of
transformations is the intermediate loss of a tilt system passing from $%
a^{-}a^{-}a^{-}$ to $a^{0}b^{-}b^{-}$. Yet, the general trend of C, R and O\
structures with decreasing $t$ is obeyed, and the final $a^{+}b^{-}b^{-}$
tilt system is the usual ground state of tilted perovskites,\cite{GB07} also
favored by the slightly covalent component of the A-O bonds.\cite%
{GK73,Woo97} It can be concluded that the tolerance factor should be the
relevant parameter in promoting the structural transformations in BCY.
Another indication in this sense is the fact that SrCeO$_{3}$, having a
still smaller $t$ due the smaller Sr ionic radius, remains in the O2 phase
at least up to 1270~K.\cite{KMB05}

It has been noted that, in perovskites with cation chemical disorder in the
A sublattice, the temperatures of the structural, and especially magnetic
and electronic transitions appear to be sensitive to both the tolerance
factor, which measures the coherent strain effect, and the variance of the A
cations sizes, which measures the incoherent part.\cite{Att01} In the
present case we are dealing only with structural transformations, without
the additional critical dependence on the bond angles involved in the
electronic and magnetic transitions, and we will just take into account the
average effects included in $t$.

For the limited objective of understanding the effect of doping and
hydration on the structural transformations, but not their detailed
nature, we keep the analysis as simple as possible, following the
idea that the driving force for the $n$-th tilting transition is the
decrease of $t$ below some critical value $t_{n}$, and therefore
that the transition temperatures are proportional to such a driving
force,
\begin{equation}
T_{n}\left( x\right) =\Delta T\times \left( t_{n}-t\left( x\right) \right) ~.
\label{Tn1}
\end{equation}%
The dependence of $t$ on $x$ can be estimated by assuming Vegard's law,
namely that the introduction of a molar concentration $x$ of defects, each
contributing with a change $\delta v$ to the ionic volume, causes an
isotropic volume change equal to $\Delta V=\delta v~x$. In addition we
consider a purely ionic picture with each ion having its nominal valence.
The ionic species and their molar fractions, coordination numbers and radii
are listed in Table I.

\begin{table}[tbp]
\caption{Ionic species of BCY and their molar fractions, radii and
coordination numbers, according to Shannon}%
\begin{ruledtabular}
\begin{tabular}{r r r r}
ion & molar fraction & radius (\AA ) & CN \\
Ba$^{2+}$ & $1$ & 1.61 & 12 \\
Ce$^{4+}$ & $1-x$ & 0.87 & 6 \\
O$^{2-}$ & $3-\delta -y$ & 1.35 & 2 \\
Y$^{3+}$ & $x$ & 0.90 & 6 \\
$\text{V}_{\text{O}}$ & $\delta $ & 1.35 &  \\
OH$^{-}$ & $y$ & 1.32 & 2%
\end{tabular}
\end{ruledtabular}
\end{table}

Notice that there is no difference between the use of ionic and crystal
radii of the Shannon tables,\cite{SP69} since they all differ by $\pm 0.14$~%
\AA , depending whether they are anions or cations, and the A-O and B-O\
ideal distances are unaffected by the choice. The tolerance factor of
undoped BaCeO$_{3}$ resulting from Table I, to be considered as referred to
the O2 room temperature structure, is $t_{0}=0.9428$. Dealing with complete
outgassing or hydration, we discard the electronic compensation and assume
that the chemical formula of BCY with $\delta $ V$_{\text{O}}$ and $\frac{y}{%
2}$H$_{2}$O is Ba$^{2+}$Ce$_{1-x}^{4+}$Y$_{x}^{3+}$O$_{3-\delta -y}^{2-}$OH$%
_{y}^{-}$, with the charge compensation requiring
\begin{equation*}
2\delta +y=x
\end{equation*}%
Hydrated and outgassed states are therefore defined by $y=x,\delta =0$ and $%
y=0,\delta =x/2$, respectively. The proton is assumed to form the
hydroxide complex (OH)$^{-}$, whose radius is also tabulated. The
assumption is corroborated by the observation that H fills the hole
at the top of the bonding Ce $4f-$O $2p$ valence band, mostly of O
$2p$ character, which is introduced by trivalent doping in dry
atmosphere.\cite{HTM05} Certainly, the approximation of a spherical
(OH)$^{-}$ ion is inadequate, but this is discussed later on.

The tolerance factor of doped BCY can then be written as
\begin{equation*}
t=\frac{1}{\sqrt{2}}\frac{\left\langle d_{\text{AO}}\right\rangle }{%
\left\langle d_{\text{BO}}\right\rangle }
\end{equation*}%
with
\begin{eqnarray*}
\left\langle d_{\text{AO}}\right\rangle &=&r_{\text{Ba}}+\left( 1-y/3\right)
r_{\text{O}}+y/3~r_{\text{OH}}=d_{\text{AO}}^{0}+\frac{y}{3}\Delta r_{\text{O%
}} \\
\left\langle d_{\text{BO}}\right\rangle &=&\left( 1-x\right) r_{\text{Ce}%
}+xr_{\text{Y}}+\left( 1-y/3\right) r_{\text{O}}+y/3~r_{\text{OH}}= \\
&=&d_{\text{BO}}^{0}+x\Delta r_{\text{B}}+\frac{y}{3}\Delta r_{\text{O}}
\end{eqnarray*}%
where $\Delta r_{\text{B}}=r_{\text{Y}}-r_{\text{Ce}}>0$, $\Delta r_{\text{O}%
}=$ $r_{\text{OH}}-r_{\text{O}}<0$\ is not very influent because it appears
in the same manner in the numerator and denominator of $t$, and the presence
of V$_{\text{O}}$ is not taken into account yet. It is sometimes assumed
that the radius of an V$_{\text{O}}$ in a perovskite is the same as that of
the O$^{2-}$\ ion,\cite{ZVP08} and the fact that O deficient perovskites
such as LaCoO$_{3-\delta }$ increase their volume with increasing $\delta $
is attributed to the enhanced radius of the reduced B cation.\cite%
{KKT02c,HSB03,ZVP08} In the present case, it is evident that a similar
assumption would not explain the marked depression of the transition
temperatures of the outgassed state with respect to the hydrated and the
undoped states. Such a depression would require an increase of $t$ in Eq. (%
\ref{Tn1}), which is not supported by any indication. The reduction of $%
T_{n} $ must arise from the elimination of the B-O-B and A-O-A
bonds, whose rigid networks compete against each other, with
expansive and compressive pressures respectively. The introduction
of V$_{\text{O}}$ therefore relaxes the driving force for the
tilting structural transformations, reducing it by an amount
$R\left( \delta \right) $. In the absence of a more detailed and
quantitative estimate of the structural relaxation introduced by V$_{\text{O}%
}$, we will assume that $R\left( \delta \right) =\left( 1-f~\delta \right) $%
, resulting in
\begin{equation}
T_{n}=\Delta T\times \left( t_{n}-t\right) \left( 1-f~\delta \right) ~,
\label{Tn2}
\end{equation}%
where $f$ is a parameter that quantifies the amount of lattice relaxation
associated with V$_{\text{O}}$; $f=1/3$ would correspond to a situation in
which a tilting driving force exists even with few sparse bonds, and
therefore it must be $f\gg 1/3$. The value of $f$, or more properly the
shape of the function $R\left( \delta \right) $, and particularly the value
of $\delta $ at which it vanishes, should be connected with the critical
concentration of V$_{\text{O}}$ at which continuum sequences of bonds
disappear over some length scale. Yet, there are many factors involved, for
example at $\delta =0.5$ there might be ordering into the brownmillerite
structure,\cite{SCW92,ZS95} and vanishing of $R\left( \delta \right) $ at $%
\delta =0.5$ would correspond to $f=2$. We will leave $f$ as a free
parameter whose value must be $\gg 1/3$.

From Eq. (\ref{Tn2}) we may analyze the various factors producing a
variation of $T_{n}$ with doping and hydration. In the hydrated state $R=1$
and the relevant quantity is, to first order in the changes of the ionic
radii,
\begin{equation}
\frac{d\left( t_{n}-t\right) }{dx}\simeq \left[ \frac{\Delta r_{\text{B}}}{%
d_{\text{BO}}^{0}}+\frac{\Delta r_{\text{O}}}{3}\left( \frac{1}{d_{\text{BO}%
}^{0}}-\frac{1}{d_{\text{AO}}^{0}}\right) \right] t_{0}  \label{ddtdx}
\end{equation}%
where the first positive term, representing the average increase of the B
radius on doping, is dominant. The second term is reduced by a geometrical
factor $\simeq \frac{1}{3}\left( 1-1/\sqrt{2}\right) \simeq 0.1$, and
therefore the most questionable assumption of adopting the tabulated radius
for the hydroxide ion is not important. From Eqs. (\ref{Tn2})\ and (\ref%
{ddtdx})\ we obtain the proportionality factor between tilting driving force
and structural transition temperature, as
\begin{equation*}
\Delta T=\frac{\left. \frac{dT_{n}}{dx}\right\vert _{\text{hydr}}}{\left.
\frac{d\left( t_{n}-t\right) }{dx}\right\vert _{\text{hydr}}}~.
\end{equation*}%
Setting $\left. \frac{dT_{2}}{dx}\right\vert _{\text{hydr}}=520$~K we obtain
$\Delta T=44300$~K and extract the critical tolerance factor $t_{2}$ for the
O1-R transition at $T_{2}=660$~K in the undoped case from Eq. (\ref{Tn2}) as
$t_{2}=0.9577$. Finally, the parameter $f$ is deduced from the initial slope
of $\left. \frac{dT_{2}}{dx}\right\vert _{\text{outg}}\simeq -1100$~K as
\begin{equation*}
f=\frac{2}{\left( t_{2}-t_{0}\right) }\left[ \frac{\Delta r_{\text{B}}}{d_{%
\text{BO}}^{0}}t_{0}-\frac{\left. \frac{dT_{2}}{dx}\right\vert _{\text{outg}}%
}{\Delta T}\right] =5.0~.
\end{equation*}%
The resulting $T_{2}\left( x\right) $ curves in the hydrated and outgassed
states are plotted as thick solid lines in Fig. \ref{fig5}. There is some
arbitrariness in the choice of the initial slopes of the curves, due to the
above mentioned anomaly between $x=0.1$ and 0.15, but the main features can
be reproduced with reasonable parameters. This simple reasoning might be
applied also to the other two transitions at $T_{1}$ and $T_{3}$, with
different values for the critical tolerance factors $t_{1}>$ $t_{2}>$ $t_{3}$
and possibly also for the proportionality factor $\Delta T,$ because the
different structures relax the mismatch between A-O and B-O sublattices at
varying degrees. The dashed lines in Fig. \ref{fig5} are obtained letting $%
\Delta T$ and $f$ unchanged and setting $t_{1}=0.9701$ in order to
reproduce $T_{1}\left( 0\right) =1210$~K. It is reassuring to find
that $t_{1}$ is exactly at the lower limit of the range $0.97<t<1$
where cubic perovskites are found.\cite{LBW06} There are no data for
$T_{1}$ in the hydrated state, while for the outgassed state there
are only preliminary anelastic spectra suggesting that at $x=0.15$
it is $T_{1}\gtrsim 950$~K, about 100~K higher than the dashed line.
We refrain from speculating whether this would be due to a larger
value of $\Delta T$ for that transition or to other reasons that are
not included in the present minimal model.

Additional factors are likely present in the O1-O2 transition at $T_{3}$,
whose temperature even decreases slightly on doping, maintaining the anomaly
between $x=0.10$ and 0.15. Yet, the effect of V$_{\text{O}}$ is again to
depress the transition temperature, although this is verifiable only at $%
x=0.02$, due to the overlapping with the O1-R transformation. Among the
phenomena interfering with the O1-O2 transition is the ordering of the V$_{%
\text{O}}$. In fact, while the V$_{\text{O}}$ are disordered in their three
equivalent O sublattices in the rhombohedral structure and also in the
orthorhombic O1, they are confined to only two sublattices, avoiding the
third crystallographically inequivalent sublattice, in the O2 structure.\cite%
{TLR00,Kni01} It is not clear whether V$_{\text{O}}$ ordering is
concomitant with the transition or it occurs at a slower rate after
the transition is completed. We have already noted\cite{137} that a
possible sign of cooperative ordering of V$_{\text{O}}$ is the
anelastic relaxation process labeled as R6, whose intensity seems to
diverge on approaching $T_{2}$ from below, as expected from critical
ordering of the elastic quadrupoles associated with the
V$_{\text{O}}$.\cite{BSW94,104,Cor10} A\ similar divergence of the
relaxation strength $\Delta $ is shown in Fig. \ref{fig4} for peak
RV. In the framework of the Bragg-Williams approximation,\cite{ST55}
the critical temperature for the onset of ordering of V$_{\text{O}}$
is expected to scale as\cite{BSW94,Cor10} $\delta \left( 1-\delta
\right) $ and therefore to increase with doping, possibly driving
the O2-O1 transformation
to higher temperature and explaining the apparently reduced decrease of $%
T_{3}\left( x,\delta =x/2\right) $ with respect to $T_{3}\left( x,\delta
=0\right) $, compared to $T_{2}$.

The fact that the decrease of $T_{3}$ in the hydrated state is smaller then
for $T_{2}$, instead, must involve completely different mechanisms. A
possibility is that on cooling the proton localizes itself more and more
within the plane perpendicular to the B-O-B bond,\cite{TLR00,Kni01,AI09}
effectively resulting in an increased flattening of the hydroxide ion and
hence in a reduction of its effective radius along the B-O-B bond direction.
Such an effect, namely an additional reduction of the B-OH-B but not of the
A-OH-A bond lengths on cooling, would reduce the mismatch between the two
bond networks and result in a stabilization of the higher temperature phase,
hence a decrease of $T_{3}$.

Finally, let us compare these dependencies of the transition temperatures on
Y doping with those measured by XRD and dilatometry with Yb doping.\cite%
{YY03} Those data have been considered insufficient,\cite{OHS09} due to the
smallness of the anomalies in the linear expansion, which also exhibit an
additional dip not associated with any phase transformation, and the limited
number of diffraction peaks that were analyzed. Yet, Yamaguchi and Yamada%
\cite{YY03} plotted $T_{1}$, $T_{2}$ and $T_{3}$ versus Yb doping
measured both under wet and dry conditions, as in our Fig.
\ref{fig5}. Similarly to the present results, the $T_{n}$ in dry
atmosphere are lower than those under in wet atmosphere, especially
for $x\geq 0.1$ but to a lesser extent than in Fig. \ref{fig5}. The
main difference between the two sets of experiments is that
$T_{2}\left( x\right) $ with Yb\ has a much weaker rise with doping
than with Y and only for\emph{\ }$x\geq 0.1$, while $T_{1}\left(
x\right) $ even decreases with doping. This difference can also be
explained within the above model, since the radius of Yb$^{3+}$ is
slightly smaller than that of Ce$^{4+}$, instead of larger as for
Y$^{3+}$ , so that with Yb it is $\Delta r_{\mathrm{B}}=$
$-0.003$~\AA\ instead of $+0.03$~\AA .

\section{Conclusions}

The temperatures $T_{n}$ of the structural transformations in BCY have been
systematically deduced from the anelastic spectra as a function of Y doping $%
x$ and hydration level $y$, or O\ deficiency $\delta $. The most complete
data are for the intermediate transition between orthorhombic \textit{Imma}
and rhombohedral at $T_{2}$; the data of the transition at the lowest
temperature $T_{3}$ to the orthorhombic \textit{Pnma} phase are incomplete
in the outgassed state, due to overlapping with the transition at $T_{2}$,
whose effects prevail in the anelastic spectra. Of the transition to the
cubic phase we could measure only the temperature $T_{1}=1210$~K in the
undoped state.

The main result is that $T_{2}\left( x\right) $ increases roughly as $\left(
500~\text{K}\right) \times x$ in the fully hydrated state and decreases
twice as much in the fully outgassed state. An anomaly with respect to the
monotonic trend between $x=0.10$ and 0.15 is associated with similar
anomalies already observed in various structural parameters at the same
doping level, while the average trend is explained with a simple model. As
usual, it is assumed that the main driving force for the structural
transformations, which consist of rotations of the BO$_{6}$ octahedra (B =
Ce,Y), is the mismatch between the more rigid and compressed network of
B-O-B\ bonds and the network of Ba-O-Ba bonds under expansion. The
transition temperatures $T_{n}$ are assumed to be proportional to $\left(
t_{n}-t\right) \left( 1-f~\delta \right) $, where $t$ is the tolerance
factor measuring the ratio between the ideal Ba-O and B-O bond lengths, $%
t_{n}$ a critical value of $t$ below which the $n$-th tilting transition
occurs, and $f$ a parameter determined by how much the mismatch stress
between different types of bonds is relieved by the presence of O vacancies.
In this manner it is possible to explain the experimental $T_{2}\left(
x,\delta \right) $ data with reasonable parameters, and also to reproduce $%
T_{1}\left( x=0\right) $ with $t_{1}=$ 0.97, which is just the lower limit
of the known range $0.97<t<1$ for cubic perovskites.

The transition temperature $T_{3}$, instead, decreases with doping and has a
reduced difference between hydrated and outgassed states. These differences
with respect to $T_{2}\left( x,\delta \right) $ are tentatively explained in
terms of a reduction of the mismatch between the bond lengths, due to a
flattening of the effective shape of the hydroxide ion perpendicularly to
the B-O-B bond during cooling, and to the ordering of the O\ vacancies in
the \textit{Pnma} phase.

\section*{Acknowledgments}

We wish to thank F. Corvasce, M. Latino, A. Morbidini for their technical
assistance, and Ing. E. Verona and coworkers of CNR-IDASC for the Pt
depositions. This research is supported by the FISR Project of Italian MIUR:
"Celle a combustibile ad elettroliti polimerici e ceramici: dimostrazione di
sistemi e sviluppo di nuovi materiali".\pagebreak


\begin{references}

\bibitem{Kni01}
K.S. Knight, Solid State Ion. {\bf 145}, 275-294 (2001).

\bibitem{OHS09}
T. Ohzeki, S. Hasegawa, M. Shimizu and T. Hashimoto, Solid State Ion. {\bf 180},
 1034-1039 (2009).

\bibitem{Dar96}
C.N.W. Darlington, phys. stat. sol. (a) {\bf 155}, 31 (1996).

\bibitem{Kre97}
K.D. Kreuer, Solid State Ion. {\bf 97}, 1-15 (1997).

\bibitem{BHG02}
P. Berastegui, S. Hull, F.J. Garcia-Garcia and S.-G. Eriksson, J. Solid State
Chem. {\bf 164}, 119 (2002).

\bibitem{NIS07}
T. Nagai, W. Ito and T. Sakon, Solid State Ion. {\bf 177}, 3433 (2007).

\bibitem{CMT09}
G. Chiodelli, L. Malavasi, C. Tealdi, S. Barison, M. Battagliarin, L. Doubova,
M. Fabrizio, C. Mortalo and R. Gerbasi, J. Alloys and Compounds {\bf 470}, 477 (2009).

\bibitem{BEK98}
Yu.M. Baikov, V.M. Egorov, N.F. Kartenko, B.A-T. Melekh, Yu.P. Stepanov and
Yu.N. Filin, Techn. Phys. Lett. {\bf 24}, 782 (1998).

\bibitem{YY03}
S. Yamaguchi and N. Yamada,, Solid State Ion. {\bf 162-163}, 23
(2003).

\bibitem{SVA93}
T. Scherban, R. Villeneuve, L. Abello and G. Lucazeau, Solid State
Ion. {\bf 61}, 93 (1993).

\bibitem{Kni99}
K.S. Knight, Solid State Commun. {\bf 112}, 73 (1999).

\bibitem{TLR00}
K. Takeuchi, C.-K. Loong, J.W. Richardson Jr, J. Guan, S.E. Dorris
and U. Balachandran, Solid State Ion. {\bf 138}, 63 (2000).

\bibitem{KMJ05}
A. Kruth, G.C. Mather, J.R. Jurado and J.T.S. Irvine, Solid State
Ion. {\bf 176}, 703 (2005).

\bibitem{GLD07}
F. Giannici, A. Longo, F. Deganello, A. Balerna, A.S. Arico and A.
Martorana, Solid State Ion. {\bf 178}, 587 (2007).

\bibitem{137}
F. Cordero, F. Trequattrini, F. Deganello, V. La Parola, E. Roncari and A.
Sanson, Appl. Phys. Lett. {\bf 94}, 181905 (2009).

\bibitem{132}
F. Cordero, F. Craciun, F. Deganello, V. La Parola, E. Roncari and A. Sanson,
Phys. Rev. B {\bf 78}, 054108 (2008).

\bibitem{DMD08}
F. Deganello, G. Marc{\`{\i}} and G. Deganello, J. Europ. Ceram. Soc. {\bf },
(2008).

\bibitem{NB72}
A.S. Nowick and B.S. Berry, {\it Anelastic Relaxation in Crystalline Solids}.
(Academic Press, New York, 1972).

\bibitem{Goo04}
J.B. Goodenough, Rep. Prog. Phys. {\bf 67}, 1915 (2004).

\bibitem{SP69}
R.D.Shannon and C.T. Prewitt, Acta Crystallogr., Sect. B: Struct. Sci. {\bf 25},
925 (1969).

\bibitem{Att01}
J.P. Attfield, Int. J. Inorg. Chem {\bf 3}, 1147 (2001).

\bibitem{Gla72}
A.M. Glazer, Acta Cryst. B {\bf 28}, 3384 (1972).

\bibitem{GB07}
P. Goudochnikov and A.J. Bell, J. Phys.: Condens. Matter {\bf 19}, 176201 (2007).

\bibitem{GK73}
J.B. Goodenough and J.A. Kafalas, J. Solid State Chem. {\bf 6}, 493 (1973).

\bibitem{Woo97}
P.M. Woodward, Acta Crystallogr., Sect. B: Struct. Sci. {\bf 53}, 32 (1997).

\bibitem{KMB05}
K.S. Knight, W.G. Marshall, N. Bonanos and D.J. Francis, J. Alloys
and Compounds {\bf 394}, 131 (2005).

\bibitem{HTM05}
T. Higuchi, T. Tsukamoto, H. Matsumoto, T. Shimura, K. Yashiro, T.
Kawada, J. Mizusaki, S. Shin and T. Hattori, Solid State Ion. {\bf
176}, 2967 (2005).

\bibitem{ZVP08}
A.Yu. Zuev, A.I. Vylkov, A.N. Petrov and D.S. Tsvetkov, Solid State
Ion. {\bf 179}, 1876 (2008).

\bibitem{HSB03}
K. Hilpert, R.W. Steinbrech, F. Boroom, E. Wessel, F. Meschke, A.
Zuev, O. Teller, H. Nickel and L. Singheiser, J. Eur. Ceram. Soc.
{\bf 23}, 3009 (2003).

\bibitem{KKT02c}
V.V. Kharton, A.V. Kovalevsky, E.V.TsipisA.P. Viskup, E.N.
Naumovich, J.R. Jurado and J.R. Frade, J. Solid State Electrochem.
{\bf 7}, 30 (2002).

\bibitem{SCW92}
A.F. Sammells, R.L. Cook, J.H. White, J.J. Osborne and R.C. MacDuff, Solid
State Ion. {\bf 52}, 111 (1992).

\bibitem{ZS95}
G.B. Zhang and D.M. Smyth, Solid State Ion. {\bf 82}, 161 (1995).

\bibitem{LBW06}
M.W. Lufaso, P.W. Barnes and P.M. Woodward, Acta Crystallogr., Sect.
B: Struct. Sci. {\bf 62}, 397 (2006).

\bibitem{104}
F. Cordero, M. Ferretti, M.R. Cimberle and R. Masini, Phys. Rev. B {\bf 67},
144519 (2003).

\bibitem{BSW94}
F. Brenscheidt, D. Seidel and H. Wipf, J. Alloys and Compounds {\bf
211/212}, 264 (1994).

\bibitem{Cor10}
F. Cordero, {\it Anelastic Spectroscopy Studies of High-Tc: Superconductors:
       dynamics of hole stripes, oxygen atoms and octahedra}.  (Lambert
Academic Publishing, Saarbr\"ucken, Germany, 2010).

\bibitem{ST55}
T. Muto and Y. Takagi, {\it Solid State Physics}. ed. by F. Seitz
and D. Turnbull, p. 193 (Academic Press, New York, 1955).

\bibitem{AI09}
A.K. Azad and J.T.S. Irvine, Chem. Mater. {\bf 21}, 215 (2009).


\end{references}

\end{document}